\documentclass[preprint]{aastex631} 

\newcommand*\patchAmsMathEnvironmentForLineno[1]{%
\expandafter\let\csname old#1\expandafter\endcsname\csname #1\endcsname
  \expandafter\let\csname oldend#1\expandafter\endcsname\csname end#1\endcsname
  \renewenvironment{#1}%
     {\linenomath\csname old#1\endcsname}%
     {\csname oldend#1\endcsname\endlinenomath}}%
\newcommand*\patchBothAmsMathEnvironmentsForLineno[1]{%
  \patchAmsMathEnvironmentForLineno{#1}%
  \patchAmsMathEnvironmentForLineno{#1*}}%
\AtBeginDocument{%
\patchBothAmsMathEnvironmentsForLineno{equation}%
\patchBothAmsMathEnvironmentsForLineno{align}%
\patchBothAmsMathEnvironmentsForLineno{flalign}%
\patchBothAmsMathEnvironmentsForLineno{alignat}%
\patchBothAmsMathEnvironmentsForLineno{gather}%
\patchBothAmsMathEnvironmentsForLineno{multline}%
}

\usepackage{textcomp}  
\usepackage{comment}
\usepackage{subfigure}
\usepackage{physics}
\usepackage{color}
\usepackage{booktabs}   

\shorttitle{AstroNuSearchKam2026}
\shortauthors{KamLAND collaboration}

\graphicspath{{./}{figures/}}

\begin{document}

\title{Search for Diffuse Supernova Neutrino Background in the Full KamLAND Dataset with Neural-Network-Based Event Classification}


\newcommand{\tohoku}{\affiliation{Research Center for Neutrino
    Science, Tohoku University, Sendai 980-8578, Japan}}
\newcommand{\ipmu}{\affiliation{Kavli Institute for the Physics and Mathematics of the Universe (WPI), 
    The University of Tokyo Institutes for Advanced Study, 
    The University of Tokyo, Kashiwa, Chiba 277-8583, Japan}}
\newcommand{\tokyorika}{\affiliation{Department of Physics and Astronomy, Tokyo University of Science, Noda, Chiba 278-8510, Japan}}
\newcommand{\osaka}{\affiliation{Graduate School of 
    Science, Osaka University, Toyonaka, Osaka 560-0043, Japan}}
\newcommand{\rcnp}{\affiliation{Research Center for Nuclear Physics, 
    Osaka University, Ibaraki, Osaka 567-0047, Japan}}
\newcommand{\tokushima}{\affiliation{Department of Physics, 
    Tokushima University, Tokushima 770-8506, Japan}}
\newcommand{\tokushimags}{\affiliation{Graduate School of Integrated Arts and Sciences, 
    Tokushima University, Tokushima 770-8502, Japan}}
\newcommand{\nagoya}{\affiliation{Department of Energy Science and Engineering, Nagoya University, Furo, Nagoya 464-8603, Japan}}
\newcommand{\lbl}{\affiliation{Nuclear Science Division, Lawrence Berkeley National Laboratory,
    Berkeley, California 94720, USA}}
\newcommand{\hawaii}{\affiliation{Department of Physics and Astronomy,
    University of Hawaii at Manoa, Honolulu, Hawaii 96822, USA}}
\newcommand{\mitech}{\affiliation{Massachusetts Institute of Technology, 
    Cambridge, Massachusetts 02139, USA}}
\newcommand{\ut}{\affiliation{Department of Physics and
    Astronomy, University of Tennessee, Knoxville, Tennessee 37996, USA}}
\newcommand{\tunl}{\affiliation{Triangle Universities Nuclear Laboratory, Durham, 
    North Carolina 27708, USA; \\
    Physics Departments at Duke University, Durham, North Carolina 27708, USA; \\
    North Carolina Central University, Durham, North Carolina 27707, USA; \\
    and The University of North Carolina at Chapel Hill, Chapel Hill, North Carolina 27599, USA}}
\newcommand{\vt}{\affiliation{Center for Neutrino
   Physics, Virginia Polytechnic Institute and State University, Blacksburg,
   Virginia 24061, USA}}
\newcommand{\washington}{\affiliation{Center for Experimental Nuclear Physics and Astrophysics, 
    University of Washington, Seattle, Washington 98195, USA}}
\newcommand{\nikhef}{\affiliation{Nikhef and the University of Amsterdam, 
    Science Park, Amsterdam, the Netherlands}}
\newcommand{\gppu}{\affiliation{Graduate Program on Physics for the Universe, Tohoku University, Sendai 980-8578, Japan}}
\newcommand{\bu}{\affiliation{Boston University, Boston, Massachusetts 02215, USA}}
\newcommand{\chapel}{\affiliation{UNC Physics and Astronomy, 120 E. Cameron Ave., Phillips Hall CB3255, Chapel Hill, NC 27599}}
\newcommand{\obihiro}{\affiliation{Department of Human Science, Obihiro University of Agriculture and Veterinary Medicine, Obihiro, Hokkaido 080-8555, Japan}}
\newcommand{\sandiego}{\affiliation{Hal{\i}c{\i}o\u{g}lu Data Science Institute, Department of Physics, University of California San Diego, La Jolla, California, 92093, USA}}
\newcommand{\delaware}{\affiliation{Department of Physics and Astronomy, University of Delaware, Newark, Delaware 19716, USA}}

\newcommand{\aticrrnow}{\altaffiliation
    {Present address: Kamioka Observatory, Institute for Cosmic-Ray Research, 
    The University of Tokyo, Hida, Gifu 506-1205, Japan}}
\newcommand{\atrikennow}{\altaffiliation
    {Present address: Center for Advanced Photonics, 
    RIKEN, Wako, Saitama, 351-0198, Japan}}
\newcommand{\atbutsuryonow}{\altaffiliation
    {Present address: Faculty of Health Sciences, 
    Butsuryo College of Osaka, Sakai, Osaka 593-8328, Japan}}
\newcommand{\atkandainow}{\altaffiliation
    {Present address: Faculty of Environmental and Urban Engineering,
    Kansai University, Suita, Osaka 564-8680, Japan}}
\newcommand{\atkinkennow}{\altaffiliation
    {Present address: Institute for Materials Research, 
    Tohoku University, 2-1-1 Katahira, Aoba-ku, Sendai, 
    Miyagi, 980-8577, Japan}}
\newcommand{\atwakasanow}{\altaffiliation
    {Present address: Irradiation Support Division, The Wakasa Wan Energy Research Center, Tsuruga, Fukui, 914-0192, Japan}}
\newcommand{\atkurnsnow}{\altaffiliation
    {Present address: Institute for Integrated Radiation and Nuclear Science, Kyoto University, Kumatori, Osaka 590-0496, Japan}}
\newcommand{\atqstnow}{\altaffiliation
    {Present address: NanoTerasu Center, National Institutes for Quantum Science and Technology (QST), Sendai, Miyagi 980-8572, Japan}}

\author[0000-0001-6162-3453]{D.~Chernyak}\tohoku
\author{T.~Eda}\tohoku
\author[0000-0002-1043-3438]{M.~Eizuka}\tohoku 
\author{R.~Endo}\tohoku
\author[0009-0004-9302-5959]{A.~Gando}\obihiro\tohoku
\author{Y.~Gando}\obihiro\tohoku
\author[0000-0002-4238-7990]{T.~Hachiya}\tohoku
\author{F.~Haneishi}\tohoku
\author[0000-0002-3458-1144]{K.~Hata}\tohoku
\author{T.~Hirai}\tohoku
\author[0000-0002-8766-3629]{K.~Hosokawa}\tohoku
\author[0000-0001-9783-5781]{K.~Ichimura}\tohoku
\author{H.~Ikeda}\tohoku
\author{K.~Inoue}\tohoku
\author[0000-0001-9271-2301]{K.~Ishidoshiro}\tohoku
\author[0000-0002-6265-1197]{Y.~Kamei}\atrikennow\tohoku
\author[0000-0003-2350-2786]{N.~Kawada}\tohoku
\author{Y.~Kishimoto}\tohoku
\author{M.~Koga}\tohoku\ipmu
\author{K.~Mikami}\tohoku
\author[0009-0001-4510-0641]{H.~Miyake}\tohoku
\author{K.~Mizukoshi}\tohoku
\author[0009-0002-6277-4047]{D.~Morita}\tohoku
\author{K.~Nakamura}\atkandainow\tohoku
\author{R.~Nakamura}\tohoku
\author{J.~Nakane}\tohoku
\author{Y.~Nakano}\tohoku
\author{N.~Obata}\tohoku
\author{Y.~Ota}\tohoku
\author[0009-0008-8794-283X]{K.~Saito}\tohoku
\author{I.~Sakaki}\tohoku
\author[0000-0003-2705-6461]{I.~Shimizu}\tohoku
\author{J.~Shirai}\tohoku
\author{A.~Suzuki}\tohoku
\author{K.~Tachibana}\atqstnow\tohoku
\author{A.~Takeuchi}\tohoku
\author{K.~Tamae}\tohoku
\author[0000-0002-2363-5637]{H.~Watanabe}\tohoku
\author[0009-0008-6580-0259]{Z.~Xu}\tohoku

\author[0000-0002-6970-9021]{S.~Yoshida}\osaka

\author{S.~Umehara}\rcnp

\author[0000-0001-5107-6724]{K.~Fushimi}\tokushima
\author{K.~Kotera}\atwakasanow\tokushimags
\author{Y.~Urano}\atkurnsnow\tokushimags

\author{S.~Kurosawa}\nagoya

\author[0000-0003-3682-432X]{H.~Ozaki}\tokyorika

\author{B.E.~Berger}\lbl
\author[0000-0002-7001-717X]{B.K.~Fujikawa}\ipmu\lbl

\author{J.G.~Learned}\hawaii
\author{J.~Maricic}\hawaii
\author{Z.~Li}\hawaii

\author{Z.~Fu}\mitech
\author{S.~Ghosh}\mitech
\author[0000-0002-1916-4182]{J.~Smolsky}\mitech
\author{L.A.~Winslow}\mitech

\author[0000-0002-5132-3112]{Y.~Efremenko}\ipmu\ut

\author[0000-0003-0734-655X]{H.J.~Karwowski}\tunl
\author[0000-0002-2313-5763]{D.M.~Markoff}\tunl
\author[0000-0003-4031-6926]{W.~Tornow}\ipmu\tunl

\author{S.~Dell'Oro}\vt
\author{T.~O'Donnell}\vt

\author[0000-0002-9050-4610]{J.A.~Detwiler}\ipmu\washington
\author{S.~Enomoto}\washington\tohoku

\author[0000-0002-1577-6229]{M.P.~Decowski}\nikhef
\author[0000-0003-4243-4862]{K.M.~Weerman}\nikhef

\author{C.~Grant}\bu
\author{\"{O}.~Penek}\bu
\author{H.~Song}\bu

\author{A.~Li}\sandiego

\author[0000-0001-8866-3826]{S.N.~Axani}\delaware
\author{M.~Garcia}\delaware
\author[0009-0000-5440-5773]{M.~Sarfraz}\delaware
\author{C.~Lin}\delaware


\collaboration{74}{(KamLAND Collaboration)}

\begin{abstract}
We report a search for the diffuse supernova neutrino background (DSNB) with the KamLAND detector, targeting electron antineutrinos ($\bar{\nu}_e$) via inverse beta decay in the neutrino energy range of 8.3--30.8\,MeV. Using an exposure of 9.02\,kton-year (8.3--9.3\,MeV) and 9.42\,kton-year (9.3--30.8\,MeV) of liquid scintillator, we observe seven candidate events after applying a new deep-neural-network-based event classification technique. This is consistent with the background-only expectation of $16.2 \pm 9.4$ events, which includes systematic uncertainties associated with the neural-network selection. A spectral analysis of the energy and radial distributions finds no significant excess attributable to the DSNB. We therefore set 90\% confidence-level upper limits on the DSNB flux of 38--43\,$\mathrm{cm^{-2}\,s^{-1}}$, depending on the assumed DSNB model. We also derive model-independent 90\% confidence-level upper limits on the $\bar{\nu}_e$ flux, which are among the most stringent constraints below 13.3\,MeV. Beyond the DSNB search itself, this work establishes neural-network-based event classification as a promising approach for suppressing neutron-associated backgrounds in liquid-scintillator neutrino detectors. The improved sensitivity also enables an astrophysical interpretation of the average neutrino emission from core-collapse supernovae and highlights the complementary role of low-threshold liquid-scintillator detectors in future DSNB searches.
\end{abstract}


\section{Introduction}\label{sec:intro}
Core-collapse supernovae mark the final stage of stellar evolution, and 
unveiling their explosion mechanism requires direct information from the stellar core. Neutrinos emitted during the explosion provide such a probe. The vast majority of the gravitational binding energy released in a supernova is carried away by neutrinos, and because of their weakly interacting nature, they escape from deep inside the star while retaining information about the conditions at production. Measurements of the total neutrino yield, energy spectrum, and time profile therefore offer a unique window into the supernova explosion. Motivated by this, many neutrino detectors have searched for supernova neutrinos~\citep{Ikeda_2007, Mori_2022, Abe_2022, Agafonova_2015, NOVOSELTSEV2020102404, Aharmim_2011, PhysRevD.81.032001, 1994ApJ...428..629M}. However, present detectors are sensitive primarily to supernovae in the Milky Way or nearby galaxies such as the Large Magellanic Cloud, and to date the only observational example remains SN 1987A~\citep{PhysRevLett.58.1490, PhysRevD.38.448, PhysRevLett.58.1494, PhysRevD.37.3361, ALEXEYEV1988209}. Because such events are rare, direct observations of supernova neutrinos are episodic and their statistics depend strongly on chance occurrences of nearby progenitors.

An alternative and complementary target is the diffuse supernova neutrino background (DSNB), also referred to as supernova relic neutrinos (SRN). The DSNB is defined as the redshift-integrated flux of neutrinos emitted by all past core-collapse supernovae throughout cosmic history, weighted by the supernova rate. Unlike a single nearby burst, the DSNB encodes an average neutrino spectrum, enabling constraints on typical supernova properties such as the neutrino temperature, the nuclear equation of state, and the fraction of core collapses that form black holes. The DSNB may also provide independent information on the cosmic star-formation history. Supernova rates are commonly inferred from the star-formation rate (SFR) and the initial mass function (IMF), where the SFR is primarily determined from electromagnetic observations. Such measurements can miss failed supernovae with little electromagnetic emission and are subject to systematic effects such as dust attenuation. Neutrino observations therefore offer a complementary way to test the SFR independently of electromagnetic surveys.


Searches for the DSNB have been conducted with several neutrino detectors. Specifically, one of the most stringent experimental constraints to date has been obtained by the water-Cherenkov detector Super-Kamiokande. A combined analysis of data from the SK-I through SK-IV operational phases above 17.3\,MeV reported an excess at the $\sim 1.5 \sigma$ level~\citep{PhysRevD.104.122002}. In addition, subsequent SK-VI and SK-VII searches in the gadolinium-loaded phase of Super-Kamiokande (SK-Gd) reported an excess at the $\sim 1.2 \sigma$ level for energies above 9.3\,MeV~\citep{abe2025searchdiffusesupernovaneutrino}. Also, Sudbury Neutrino Observatory (SNO)~\citep{PhysRevD.70.093014} and Borexino~\citep{AGOSTINI2021102509} have reported upper limits on the model-independent antineutrino flux. These results highlight the importance of further DSNB searches with improved sensitivity and complementary detector technologies.

Kamioka Liquid Scintillator Anti-Neutrino Detector (KamLAND), a 1-kton liquid-scintillator (LS) detector, has performed DSNB searches in the neutrino energy region above 8.3\,MeV. The previous KamLAND search found a best-fit DSNB signal consistent with zero and set 90\% confidence-level (CL) upper limits of 60--110\,$\mathrm{cm^{-2}\,s^{-1}}$ on the flux, depending on the assumed DSNB model~\citep{Abe_2022}. The relatively low energy threshold of KamLAND provides sensitivity to a different part of the DSNB spectrum from that primarily probed by Super-Kamiokande. At low energies below $\sim 10\,\mathrm{MeV}$, the DSNB spectrum includes contributions from higher-redshift core-collapse supernovae at $z \sim 1\text{--}2$, whereas at higher energies around $\sim 30\,\mathrm{MeV}$ it can be particularly sensitive to hotter neutrino emission, such as that associated with black-hole formation. These considerations motivate DSNB searches over a broad energy range with multiple neutrino detectors. In KamLAND, electron antineutrinos, $\bar{\nu}_e$, are detected via inverse beta decay (IBD, $\bar{\nu}_e + p \rightarrow e^+ + n$). The positron deposits energy and annihilates with an electron, emitting two 511-keV gamma rays that constitute the prompt signal. The accompanying neutron is subsequently captured on a proton or carbon nucleus in the LS, emitting gamma rays that form the delayed signal. This delayed-coincidence (DC) signature provides powerful background suppression. However, interactions other than IBD that produce neutrons can also mimic IBD-like DC pairs. In previous KamLAND DSNB searches, neutral-current (NC) interactions induced by atmospheric neutrinos provided a particularly large background contribution and limited the sensitivity. Since the particles producing the IBD signal and neutron-associated backgrounds propagate through the detector on different time and spatial scales, their photomultiplier-tube (PMT) hit patterns contain information that can be used to distinguish these event classes. To take advantage of this information, we employ a neural-network-based event-classification method.

In this paper, we report a DSNB search using the full KamLAND data-taking period. Section~\ref{sec:detector} describes the KamLAND detector and the dataset used in this analysis. Section~\ref{sec:selection_conv} presents the conventional antineutrino selection used to identify DSNB candidates. Section~\ref{sec:selection_KamNet} describes the neural-network-based event selection developed to suppress neutron-associated backgrounds. Section~\ref{sec:background} estimates the expected event yields of each background component after applying the event selection. Section~\ref{sec:analysis} presents the DSNB search results. 
In addition to presenting the improved DSNB search results, we discuss their implications for the average neutrino emission from core-collapse supernovae in Sec.~\ref{sec:discussion}.
Section~\ref{sec:summary} summarizes our conclusions.

\section{KamLAND Detector and Dataset} \label{sec:detector}
The KamLAND detector was located 1000\,m underground at Mt. Ikenoyama in Kamioka, Japan. The corresponding overburden of 2700\,m water equivalent reduced the cosmic-ray muon flux to $\mathcal{O}(10^{-5})$ of that at the surface. The detector consisted of an inner detector (ID) housed in an 18-m-diameter stainless-steel spherical tank and an outer detector (OD) surrounding it. The ID contained 1\,kton of LS enclosed in a 13-m-diameter nylon/EVOH (ethylene-vinyl alcohol copolymer) balloon, which was supported by non-scintillating buffer oil. A total of 1325 17-inch PMTs and 554 20-inch PMTs were mounted on the inner surface of the stainless-steel tank to detect scintillation light produced in the ID. The OD was a 20-m-diameter cylindrical water-Cherenkov detector filled with 3.2\,kton of purified water; it served as a muon veto by detecting Cherenkov light from cosmic-ray muons.

KamLAND began data taking in March 2002 and ended data taking in August 2024. To remove radioactive impurities in the LS, purification campaigns were conducted in two periods: from May to August 2007 and from July 2008 to February 2009. From August 2011 to December 2015, the KamLAND-Zen 400 experiment searched for neutrinoless double-beta decay, during which a 3.08-m-diameter inner balloon filled with xenon-loaded liquid scintillator~\citep{PhysRevLett.117.082503} was installed at the center of the ID. From January to June 2016, the OD was refurbished, and 225 20-inch PMTs were replaced with 140 new 20-inch PMTs~\citep{Ozaki:2017TS}. Subsequently, from May 2018 to May 2024, the KamLAND-Zen 800 experiment~\citep{jkf6-48j8}, which employed approximately twice the xenon mass of KamLAND-Zen 400, was operated with a 3.80-m-diameter inner-balloon at the detector center. During the early phase of KamLAND-Zen 800, an increasing number of PMTs with reduced photon-detection efficiency was observed due to aging-related gain degradation. To mitigate this issue, the installation of amplifiers in the KamLAND front-end electronics began in March 2020 and was carried out across 64 separate interventions.

\section{Conventional antineutrino selection} \label{sec:selection_conv}
The positron produced in IBD interaction promptly annihilates with an electron, emitting two 511-keV gamma rays. Scintillation light from the positron kinetic energy deposition and Compton scattering of the annihilation gammas is reconstructed as a single prompt event. The neutron produced in IBD is subsequently captured on a proton (or a carbon nucleus) after a mean capture time of $207 \pm 2.8\,\mathrm{\mu s}$~\citep{PhysRevC.81.025807}. The neutron capture emits a monoenergetic gamma ray of 2.2\,MeV (4.9\,MeV) for capture on hydrogen (carbon), which is reconstructed as the delayed event. The DC signature between the prompt and delayed events provides powerful background suppression.

The DC selection requires the spatial and temporal separations between the prompt and delayed events to satisfy $\Delta R < 160\,\mathrm{cm}$ and $\Delta T = 0.5\text{--}1000\,\mathrm{\mu s}$, respectively. The energy selection criteria are based on the visible energy reconstructed from the PMT charge and timing information. The prompt visible-energy window is set to $E_{\mathrm{prompt}} = 7.5\text{--}30\,\mathrm{MeV}$, corresponding to the``golden window'' for DSNB searches between the reactor and atmospheric neutrino spectra. The delayed visible-energy window is chosen to detect the neutron-capture gamma rays: $E_{\mathrm{delayed}} = 1.8\text{--}2.6\,\mathrm{MeV}$ (4.4--5.6\,MeV) for captures on hydrogen (carbon). In the DSNB prompt-energy range, backgrounds from fast neutrons are present and increase toward the outer region of the detector. To suppress these backgrounds, we require the reconstructed vertex to be within 550\,cm from the detector center. This fiducial volume corresponds to a number of target protons of $N_p = (4.6 \pm 0.1) \times 10^{31}$. During the KamLAND-Zen 400/800 periods, an inner-balloon was installed at the detector center. Because the inner-balloon and its supporting structures can act as sources of background, we apply an additional fiducial cut during these periods that excludes (i) a spherical region with a radius of 250\,cm around the detector center and (ii) a cylindrical region of radius 250\,cm in the upper half of the detector. 
The IBD interactions are simulated using the cross section of Strumia and Vissani \citep{Strumia:2003zx}. The IBD detection efficiency for the DC selection is estimated using Monte Carlo (MC) simulations with uniformly generated neutrino events, yielding $\epsilon_{\mathrm{IBD}} = 92\%$ without, and 73\% with, the inner-balloon cut.

\section{Neural-network-based event selection}\label{sec:selection_KamNet}
The dominant background in previous KamLAND DSNB searches arises from NC interactions of atmospheric neutrinos~\citep{Abe_2022}. In these reactions, atmospheric neutrinos interact with $^{12}\mathrm{C}$ and typically produce an excited $^{11}\mathrm{C}$ nucleus accompanied by a neutron. The $^{11}\mathrm{C}$ de-excites promptly, emitting a gamma ray with the energy of 2\,MeV. The neutron ejected from $^{12}\mathrm{C}$ can have kinetic energy of $\sim 200\,\mathrm{MeV}$ and lose energy through repeated scatters on protons as it propagates through the detector. The combination of de-excitation gamma ray and recoil protons forms the prompt signal, while the subsequent neutron capture produces the delayed signal. Improving DSNB search sensitivity therefore requires a method to distinguish IBD events from neutron-associated backgrounds that include proton recoils. To this end, we exploit differences in the characteristic time and length scales of particle propagation in the prompt signal. In IBD events, the positron and annihilation gamma rays traverse distances on the order of tens of centimeters within a few nanoseconds. In contrast, proton recoils induced by neutrons occur over tens of nanoseconds, and the interaction points can extend over a spatial scale of a few meters from the neutron production point. This difference leads to distinct light-emission patterns in the detector and, consequently, differences in the PMT hit information: for events accompanied by neutron-induced proton recoils, the spatial distribution of hit PMTs evolves over a longer time scale.

\begin{figure}[htbp]
    \centering
    \includegraphics[width=0.7\linewidth]{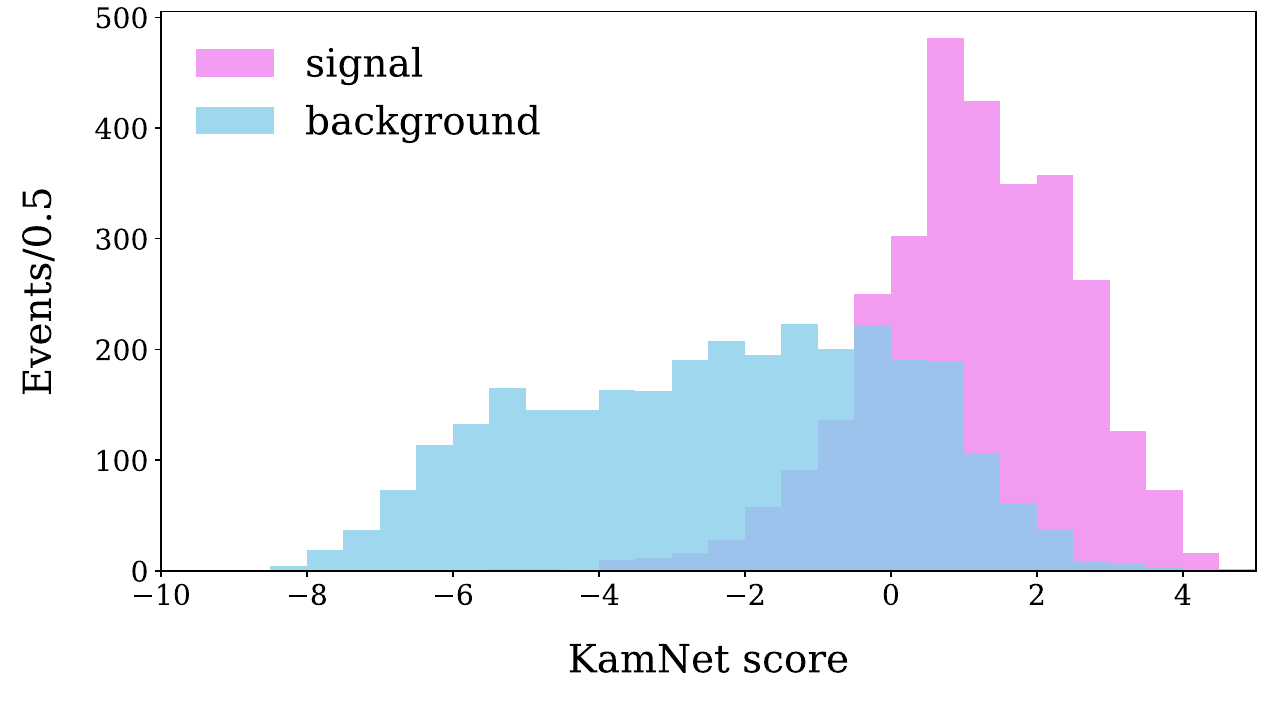}
    \caption{KamNet score distribution for the IBD signal and atmospheric-neutrino NC background. This figure corresponds to the period before purification.}
    \label{fig:KamNet_score}
\end{figure}

To utilize this information, we employ KamNet~\citep{PhysRevC.107.014323}, a deep neural network designed for KamLAND's event data structure. KamNet combines a ConvLSTM~\citep{NIPS2015_07563a3f} architecture suitable for capturing spatiotemporal features with a spherical CNN~\citep{cohen2018sphericalcnns} that preserves the detector's spherical symmetry and extracts equivariant features. KamNet was originally developed for particle classification in the KamLAND-Zen experiment. The input is a three-dimensional array binned in two spatial coordinates (PMT azimuth and zenith position) and in PMT hit time, and the output is a single score representing the signal-likeness of an event. In this work, we train KamNet using simulated PMT hit information for the prompt signals of IBD events and atmospheric-neutrino NC events. Figure~\ref{fig:KamNet_score} shows an example of the KamNet score distributions for IBD and atmospheric-neutrino NC interactions. Because the optimal KamNet threshold depends on whether reactor $\bar{\nu}_e$ are present, we divide the analysis energy region at 8.5\,MeV when applying KamNet-related selections. Additionally, after the first amplifier installation in March 2020, it became difficult to incorporate the resulting changes in the front-end electronics accurately in the detector simulation. Therefore, we do not apply the KamNet-based event selection to data collected after this date. These later data are still included in the DSNB search, but the KamNet-based selection is applied only in periods for which the detector simulation reliably reproduces the front-end response. Efficiencies and expected event yields are evaluated period by period accordingly. The KamNet acceptance for IBD-like events and rejection efficiency for atmospheric-NC-like backgrounds vary by period and energy range. The values used in the analysis are summarized in Table~\ref{tab:KamNet_eff_summary}. The dataset to which KamNet is applied is divided into five periods. We define the period before LS purification as KamNet-1, the period after purification as KamNet-2, the KamLAND-Zen 400 period as KamNet-3, the interval between Zen400 and Zen800 as KamNet-4, and the period from the start of Zen800 to the amplifier installation as KamNet-5.

\begin{table}
    \caption{Summary of IBD acceptance and atmospheric neutrino NC rejection efficiencies obtained by the KamNet selection}
    \label{tab:KamNet_eff_summary}
    \centering
    \begin{tabular}{ccccc}

        \toprule

        & \multicolumn{2}{c}{IBD signal} & \multicolumn{2}{c}{Atmospheric NC background} \\
        \cmidrule(lr{0.4em}){2-3} \cmidrule(lr{0.4em}){4-5} KamNet-period & 7.5--8.5\,MeV & 8.5--30\,MeV & 7.5--8.5\,MeV & 8.5--30\,MeV \\

        \midrule
        \midrule
        
        KamNet-1 & 85.4\% & 62.9\% & 69.0\% & 91.3\% \\
        KamNet-2 & 78.4\% & 81.1\% & 72.7\% & 74.4\% \\
        KamNet-3 & 74.5\% & 77.4\% & 74.5\% & 74.8\% \\
        KamNet-4 & 71.5\% & 74.3\% & 76.8\% & 80.8\% \\
        KamNet-5 & 68.8\% & 75.7\% & 74.3\% & 78.3\% \\

        \bottomrule
    \end{tabular}
    
\end{table}

\begin{figure}[htbp]
    \centering
    \includegraphics[width=0.8\linewidth]{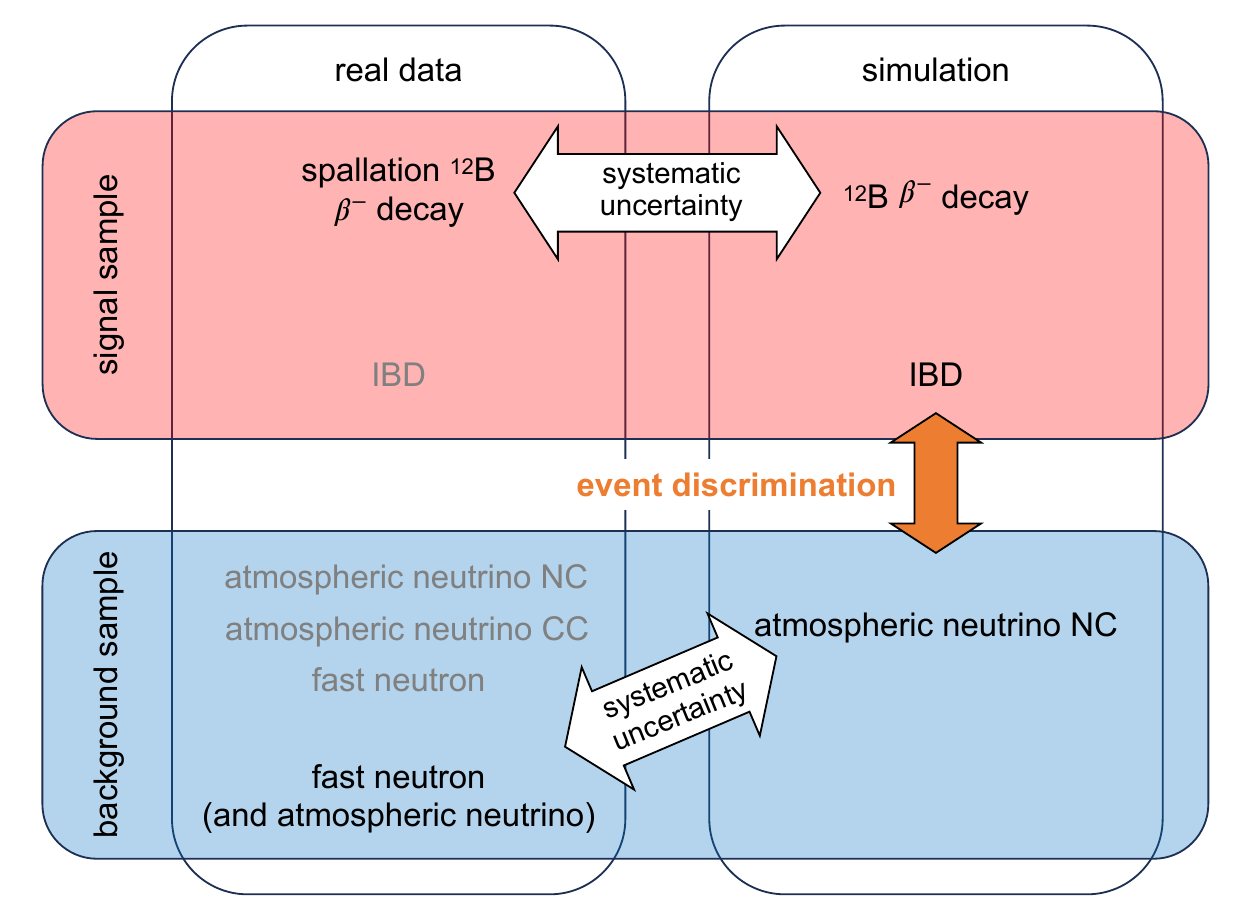}
    \caption{Schematic illustration of the proxy-sample strategy used to evaluate the systematic uncertainty associated with data–simulation differences in the KamNet selection. Simulated IBD events and atmospheric NC events are used to define the signal-like and NC-like classes, respectively. The event-discrimination arrow represents the application of the trained KamNet classifier, not the training procedure itself. Since sufficiently pure corresponding samples are not available in data, spallation $^{12}\mathrm{B}$ $\beta^-$ decays are used as an IBD-like proxy sample, while fast-neutron events are used as an NC-like proxy sample. The data–simulation differences observed in these proxy samples are propagated as systematic uncertainties on the KamNet signal acceptance and background rejection efficiency. Gray labels indicate the target event classes represented by the proxy samples.}
    \label{fig:proxy}
\end{figure}

The KamNet-related systematic uncertainty includes contributions from data-MC differences in the score distributions, statistical fluctuations in the training procedure evaluated with 100 bootstrap models, and the radial dependence of the KamNet efficiency. Differences in the KamNet score distribution between simulation and data can lead to mismodeling of the KamNet selection efficiency and must therefore be evaluated as a systematic uncertainty. Ideally, this uncertainty would be evaluated using large, pure control samples in data that closely reproduce the simulated IBD and atmospheric-neutrino NC samples. Although stopped $\mu^-$ capture provides a large sample of neutron-producing events, the neutron spectrum is dominated by relatively low-energy neutrons following nuclear capture, while also containing a higher-energy tail. Such events may provide a useful calibration sample for studying neutron-induced detector responses, and we leave a detailed investigation of their applicability to the KamNet systematic evaluation for future work. 
Instead, we estimate this systematic uncertainty by comparing data and simulation using proxy samples: $^{12}\mathrm{B}$ $\beta^-$ decays produced by muon-induced spallation are used as IBD-like samples, while fast-neutron events are used as NC-like samples with proton recoils. Although these proxy samples are not identical to the target IBD and atmospheric-neutrino NC events, they capture the relevant contrast between positron/electron-like prompt activity and proton-recoil-associated prompt activity, making them suitable proxy samples for evaluating the KamNet-related systematic uncertainty. We have verified that the difference between the KamNet selection efficiencies for IBD and $^{12}\mathrm{B}$ $\beta^-$ decays is sufficiently small for the purpose of this systematic evaluation; the same verification was performed for atmospheric-neutrino NC events and fast-neutron events. Figure~\ref{fig:proxy} illustrates the proxy-sample strategy used to estimate the systematic uncertainty arising from differences between simulation and data. To assess the statistical variability of the KamNet output, we also trained 100 bootstrap KamNet models and treated the resulting spread in the signal acceptance and background rejection efficiency as an additional systematic uncertainty. Furthermore, although KamNet is trained using events within a radius of 600,cm from the detector center in order to secure sufficient statistics, its application to the DSNB search uses events within a radius of 550,cm to avoid background events originating outside the detector. Accordingly, the radial dependence of the KamNet score is also evaluated as a source of systematic uncertainty. Combining these contributions, the overall KamNet-related systematic uncertainty is 15.0\%--39.1\% for the IBD-like signal and 36.6\%--76.3\% for proton-recoil-associated background events. The detailed values of the systematic uncertainties are summarized in Table~\ref{tab:KamNet_sys_summary}. 

\begin{table}
    \caption{Summary of KamNet-related systematic uncertainties}
    \label{tab:KamNet_sys_summary}
    \centering
    \begin{tabular}{ccccc}

        \toprule

        & \multicolumn{2}{c}{IBD signal} & \multicolumn{2}{c}{Atmospheric NC background} \\
        \cmidrule(lr{0.4em}){2-3} \cmidrule(lr{0.4em}){4-5} KamNet-period & 7.5--8.5\,MeV & 8.5--30\,MeV & 7.5--8.5\,MeV & 8.5--30\,MeV \\

        \midrule
        \midrule
        
        KamNet-1 & 35.2\% & 35.4\% & 36.6\% & 76.0\% \\
        KamNet-2 & 23.2\% & 39.1\% & 46.3\% & 76.3\% \\
        KamNet-3 & 15.5\% & 15.0\% & 42.8\% & 57.9\% \\
        KamNet-4 & 33.6\% & 35.9\% & 52.2\% & 41.9\% \\
        KamNet-5 & 38.0\% & 29.7\% & 54.3\% & 65.4\% \\

        \bottomrule
    \end{tabular}
    
\end{table}

\section{Background Estimation} \label{sec:background}
The backgrounds for the DSNB search in KamLAND can be classified according to how they mimic the DC signature of IBD. Atmospheric-neutrino interactions, fast neutrons, and spallation products can produce a prompt energy deposition accompanied by one or more neutrons, whose subsequent capture signals mimic IBD-like DC pairs, although these events are not genuine IBD interactions. Reactor antineutrinos, in contrast, interact through the same IBD channel as the DSNB signal and therefore constitute an irreducible background on an event-by-event basis. Accidental coincidences, in which two unrelated single events satisfy the DC selection criteria by chance, are also considered as a trivial but non-negligible background component.

\subsection{Atmospheric Neutrinos} \label{subsec:atmospheric}
Atmospheric-neutrino backgrounds in KamLAND include both NC interaction and charged-current (CC) interactions. In this study, we adopt the atmospheric neutrino flux calculated by \cite{PhysRevD.75.043006}. The flux uncertainty is energy dependent and is approximately 20\%. The procedure used to estimate number of atmospheric neutrino events follows that of \cite{Gando_2012}. 

At KamLAND, the dominant background from NC interactions arises from reactions in which a neutron is emitted from  a carbon nucleus, given by $\nu (\bar{\nu}) + {}^{12}\mathrm{C} \rightarrow \nu (\bar{\nu}) + n + {}^{11}\mathrm{C}^*$. In this process, the prompt signal is formed by scintillation light from recoil protons induced by the emitted neutron, together with de-excitation gamma-rays from ${}^{11}\mathrm{C}^*$. The delayed signal is subsequently provided by gamma-rays following neutron capture, and these two signals together constitute a DC signature. The cross section for NC interactions on carbon nuclei has not yet been established with sufficient precision. Therefore, we adopt the neutrino-nucleon cross section derived from the BNL measurement~\citep{PhysRevD.35.785}. An energy-dependent uncertainty of up to 18\% is assigned to these cross sections. Neutron emission from carbon nuclei can proceed from either the P-shell or the S-shell. When the neutron is emitted from the P-shell, the excited nucleus ${}^{11}\mathrm{C}^*$ de-excites by emitting a 2\,MeV gamma-ray. In contrast, when the neutron is emitted from the S-shell, the de-excitation of ${}^{11}\mathrm{C}^*$ is more complex, as described in \cite{PhysRevD.67.076007}. For each decay mode, the visible energy is evaluated by converting the particle energies with an energy-scale model that incorporates the nonlinear effects of quenching. In NC interactions, only a small fraction of the incident atmospheric neutrino energy is transferred to the neutron, resulting in neutron kinetic energies typically below 200\,MeV. Consequently, the visible-energy spectrum gradually increases toward lower energies. Then, we incorporate the KamNet rejection efficiency and the associated systematic uncertainty described in Section~\ref{sec:selection_KamNet}, and the expected number of events is estimated to be $10.1 \pm 6.2$.

In CC interactions, atmospheric neutrinos interact with either free protons or carbon nuclei. These interactions produce charged particles, such as positrons or muons, together with neutrons (and residual nuclei). The prompt event is formed by the charged particle and by proton recoils induced by neutron scattering, while the delayed event is formed by neutron capture; thus, these events can be observed through the DC signature. In the energy range relevant to this analysis, the cross section for neutrino interactions on free protons is larger than that on carbon nuclei, and therefore the former dominates the total CC contribution. Consequently, the dominant process is $\bar{\nu}_\mu + p \rightarrow \mu^+ + n$. In this study, we use the cross section based on the MiniBooNE experiment~\citep{PhysRevD.75.093003}. Muons produced in CC interactions can give rise to a threefold coincidence consisting of muon scintillation, muon decay, and neutron capture, thereby mimicking the IBD signals. Such muon-associated events can be efficiently suppressed by applying a muon veto, with an estimated rejection efficiency of 78\%~\citep{Gando_2012}. With respect to the KamNet output score, simulation-based studies indicate that the score is not significantly different from that of NC events. We therefore approximate the KamNet rejection efficiency and KamNet-related systematic uncertainty for CC events to be the same as those for NC events. With this assumption, the expected number of CC background events is estimated to be $0.5 \pm 0.3$.

\subsection{Fast Neutrons} \label{subsec:fastneutron}
High-energy neutrons can be produced when cosmic-ray muons interact with detector components or the surrounding rock. Such neutrons predominantly enter the detector from outside the ID, generate a prompt event through proton recoils, and then, after losing energy, are captured to produce a delayed event. In this way, fast-neutron events can mimic an IBD-like DC and constitute a background to the DSNB search.

The fast-neutron background rate in KamLAND was estimated in \cite{Abe_2022}. In that work, a muon simulation based on the topological map of Mt. Ikenoyama~\citep{GeographicalSurvey} and the MUSIC simulation tool~\citep{ANTONIOLI1997357} was used, and interactions in the KamLAND detector were modeled with a Geant4-based detector simulation~\citep{AGOSTINELLI2003250}. The approximate radial distribution of fast-neutron events is obtained from this simulation and is parameterized as a function of radius $R$ by $f(R) \propto \exp(R/\lambda)$, with $\lambda = 50.9 \pm 3.0\,\mathrm{cm}$. However, the absolute normalization shows a discrepancy between data and simulation. This is attributed to the limited understanding of muon-induced neutron production in the surrounding rock, and a 100\% uncertainty is therefore assigned to the predicted number of events.

Because the prompt event of fast neutron background is dominated by neutron-induced proton recoils, KamNet tends to classify them as background-like events. We estimate the KamNet rejection efficiency for fast neutrons using a fast-neutron-enriched sample selected from data with DC selection criteria, obtaining values of 71.2--93.9\%. Based on these results, the expected number of fast-neutron background events is estimated to be $2.5 \pm 2.5$ events.

\subsection{Spallation Products} \label{subsec:spallation}
Cosmic-ray muons can induce $^{12}\mathrm{C}$ spallation in the KamLAND detector, either directly or via secondary particles from hadronic interactions, producing unstable isotopes. Among these, the decay of $^{9}\mathrm{Li}$ ($\tau = 257.2\,\mathrm{ms}$ and $Q = 13.6\,\mathrm{MeV}$) is accompanied by a neutron. Consequently, the electron forms the prompt event and the neutron capture forms the delayed event, yielding an IBD-like DC and thus a background to the DSNB search.

Such $^{9}\mathrm{Li}$ spallation events can be selectively removed by constructing a likelihood function (the ``shower likelihood cut'') using the muon energy loss along its track, $dE/dx$; the distance between the muon track and the reconstructed $^{9}\mathrm{Li}$ candidate vertex; and the time difference between the muon and the $^{9}\mathrm{Li}$ event. Because the shower likelihood cut for the DSNB search was established in the previous KamLAND study~\citep{Abe_2022}, we adopt the same procedure here, except that we divide the analysis into two regions separated at 8.5\,MeV to account for the presence or absence of reactor antineutrinos. The signal inefficiency introduced by the shower likelihood cut is incorporated as a reduction in livetime; therefore, two different livetime values are used in this analysis. In this work, the spallation rejection efficiency of the shower likelihood cut is 97.6--99.7\%, and the corresponding signal inefficiency is 6.0--15.6\%, depending on the energy region and data period.

We also estimate the KamNet selection efficiency for the spallation background. To avoid reusing events that enter the final sample, we use a dataset tagged (vetoed) by the shower likelihood cut for this evaluation. We compute the KamNet score from the PMT hit information for each event and estimate the selection efficiency, obtaining values of 10.3--53.0\%. Based on these efficiencies, the expected number of spallation background events is estimated to be $1.9 \pm 6.6$ events. The uncertainty is large because, to obtain sufficient statistics for constructing the shower likelihood probability density functions, we must use $^{12}\mathrm{B}$ rather than the target $^{9}\mathrm{Li}$ sample; we therefore include as a systematic uncertainty the difference in rejection efficiency arising from the different isotope species.

\subsection{Reactor Antineutrinos} \label{subsec:reactor}
KamLAND was constructed at an average baseline of approximately 180\,km from Japanese nuclear power reactors. Following the 2011 Great East Japan Earthquake, most domestic reactors were shut down, while in recent years an increasing number of reactors have resumed operation. Reactor antineutrinos are produced primarily by the beta decays of neutron-rich fission fragments from $^{235}\mathrm{U}$, $^{238}\mathrm{U}$, $^{239}\mathrm{Pu}$, and $^{241}\mathrm{Pu}$. In the KamLAND analysis, the fission rates are calculated using the thermal power and operational status (on/off) of individual reactors.

A KamLAND measurement of reactor antineutrinos below 8.5\,MeV has been reported in \cite{PhysRevD.88.033001}; however, there is no corresponding measurement in the energy region of $\sim 10\,\mathrm{MeV}$ which is the high-energy tail of the reactor spectrum. We therefore constructed a reactor antineutrino spectrum in this energy region by adopting a polynomial function as described in \cite{Abe_2022}. Using the Huber/Mueller spectrum model~\citep{PhysRevC.84.024617, PhysRevC.83.054615} in this energy range leads to a large shape uncertainty of about 50\%. After applying the KamNet selection efficiency for IBD and combining reactor-related, detector-related, and KamNet-related uncertainties, the expected number of reactor antineutrino events is estimated to be $1.0 \pm 0.6$.

\subsection{Accidental Coincidence} \label{subsec:accidental}
Uncorrelated pairs of single events can fall within the DC window and thus constitute an accidental-coincidence background. The prompt events are dominated by radioactive impurities and long-lived spallation products, while the delayed events are primarily due to radioactive decays such as $^{208}\mathrm{Tl}$ ($2.62\,\mathrm{MeV}$ gamma ray). We estimate the accidental-coincidence rate using an off-time window from 0.01 to 20\,s after each prompt event. This off-time window is $2 \times 10^4$ times wider than the on-time window, thereby reducing the statistical uncertainty of the estimate. Because the prompt component of accidental coincidences is produced by light particles, similarly to IBD events and $^{12}\mathrm{B}$ $\beta^-$ decays, we apply the same KamNet selection efficiency and KamNet-related systematic uncertainty as used for reactor antineutrino (IBD) events. The expected number of accidental-coincidence background events is $(1.1 \pm 0.2) \times 10^{-1}$ events.

\section{Analysis and Results} \label{sec:analysis}
\begin{figure}[htbp]
    \begin{tabular}{rl}

    \begin{minipage}{0.45\columnwidth}
    \centering
        \includegraphics[width=8.5cm]{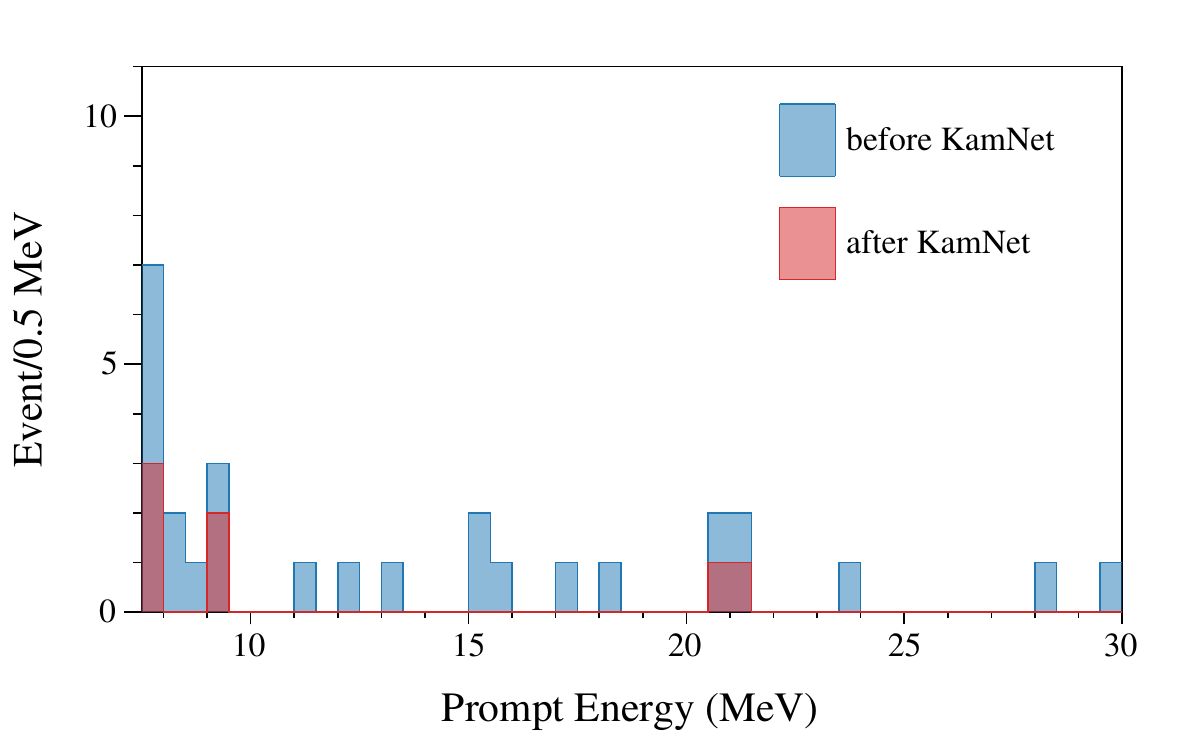}
    \end{minipage} &

    \begin{minipage}{0.45\columnwidth}
    \centering
        \includegraphics[width=8.5cm]{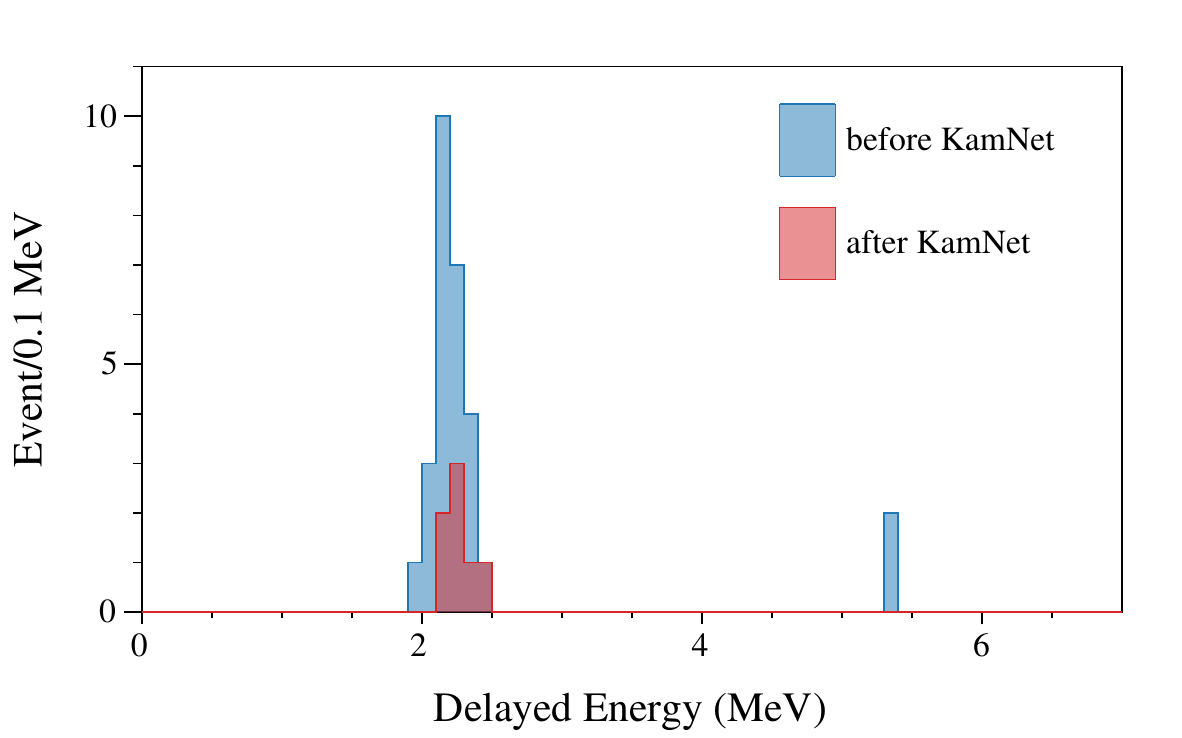}
    \end{minipage} \\
    
    \begin{minipage}{0.45\columnwidth}
    \centering
        \includegraphics[width=8.5cm]{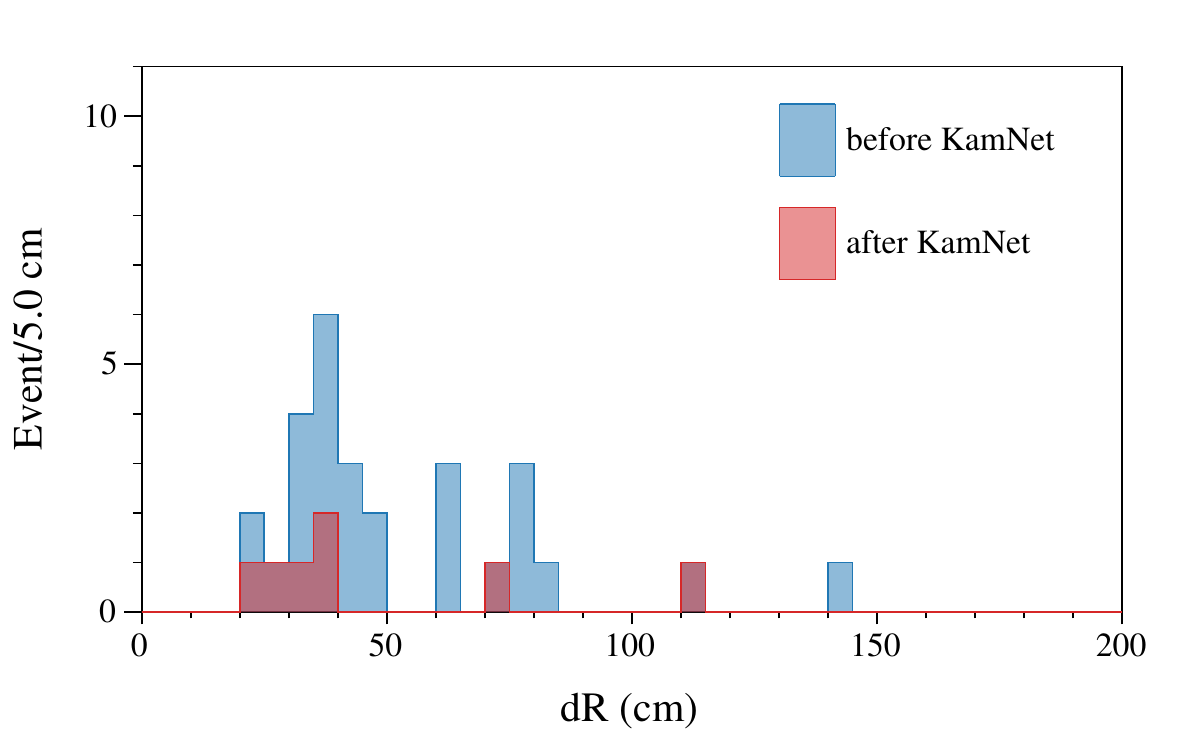}
    \end{minipage} &

    \begin{minipage}{0.45\columnwidth}
    \centering
        \includegraphics[width=8.5cm]{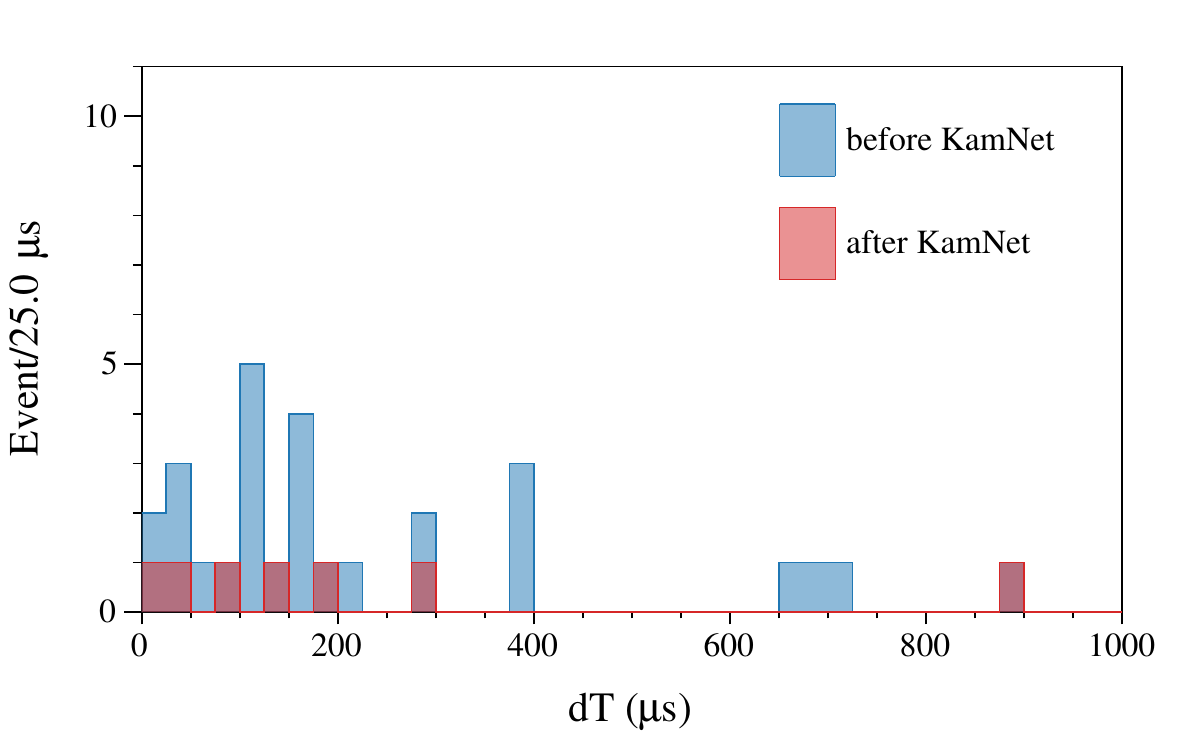}
    \end{minipage}
    
    \end{tabular}
    \caption{
    Event distributions after all conventional antineutrino selections and before (blue) and after (red) the KamNet selection;
    lower left: distance between the prompt and delayed events; lower right: time difference between the prompt and delayed events. 
    The final sample contains seven DSNB candidates.}
    \label{fig:property}
\end{figure}

\begin{figure}[htbp]
\centering
    \begin{minipage}{0.49\columnwidth}
        \centering
        \includegraphics[width=8.5cm]{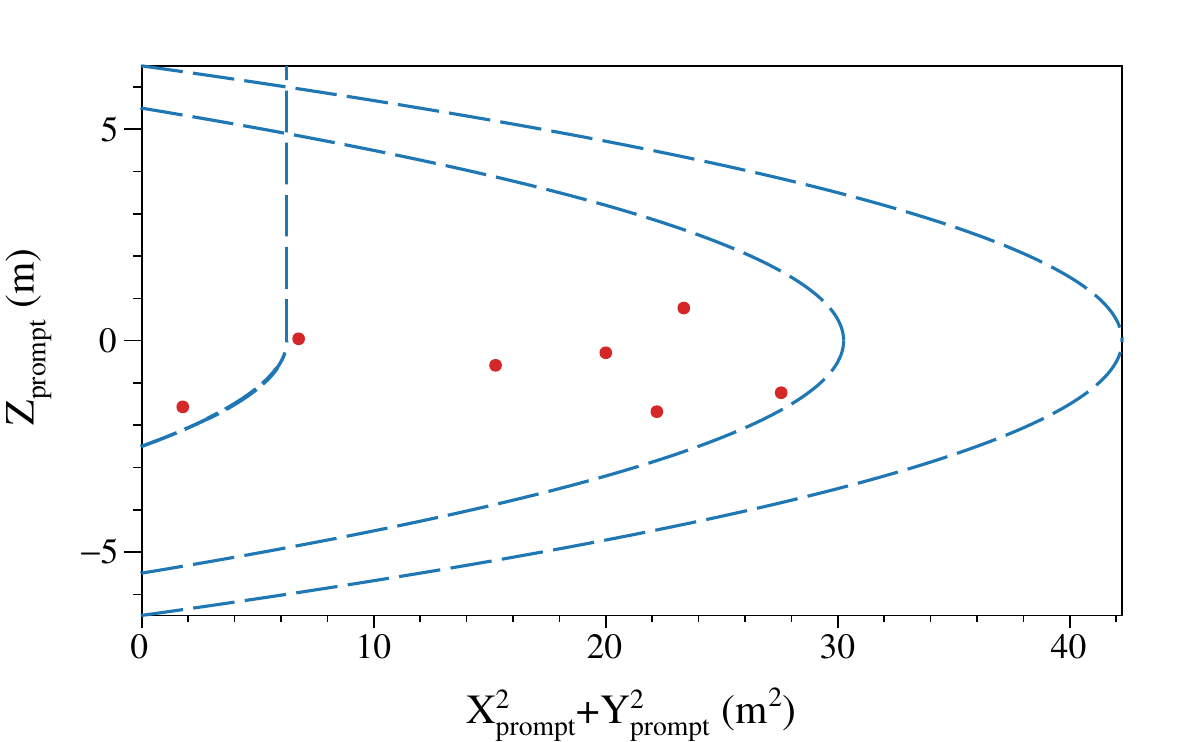}
    \end{minipage}
    \begin{minipage}{0.49\columnwidth}
        \centering
        \includegraphics[width=8.5cm]{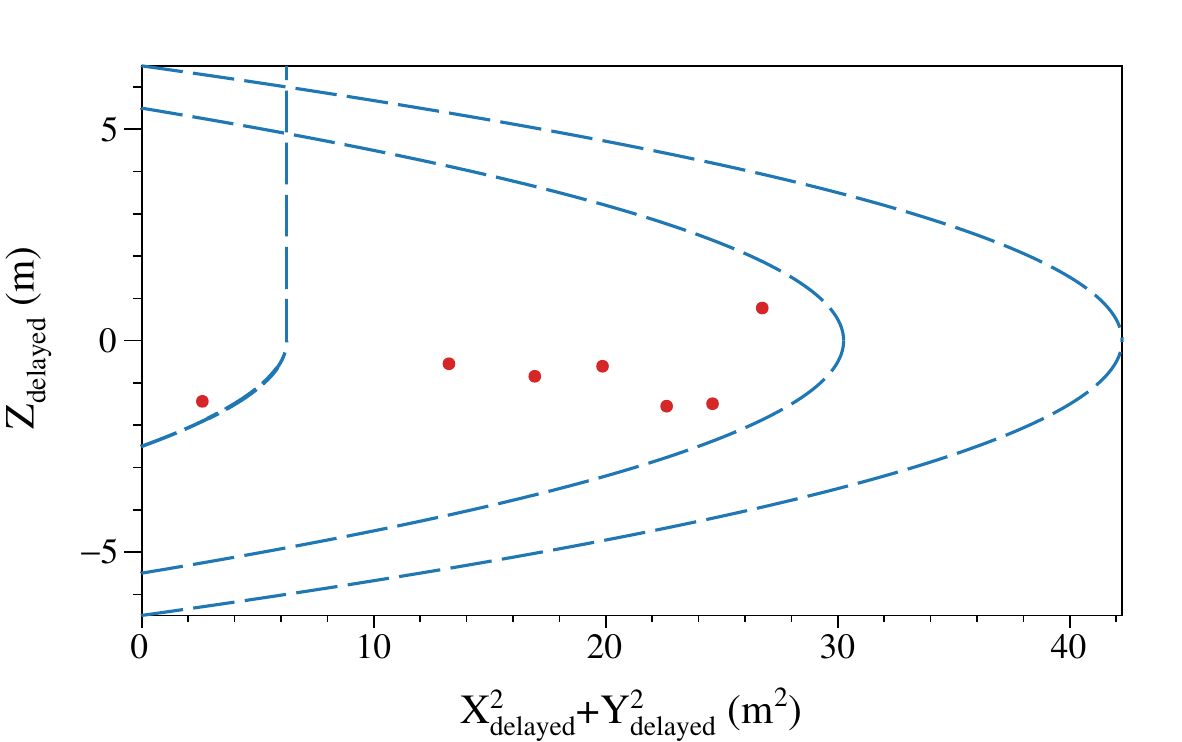}
    \end{minipage}
    \caption{Position distributions of the final DSNB candidates. The left panel shows prompt event positions while the right panel shows delayed event positions. Blue curves indicate, respectively, the inner-balloon region, whose interior is excluded from the analysis, the 550\,cm radius, and the 650\,cm radius.}
    \label{fig:property_vertex}
\end{figure}

In this paper, we perform a DSNB search using the full KamLAND dataset collected from March 2002 to August 2024 using the IBD as the detection channel. The total livetime of the search corresponds to 6082.42 days for $E_{\mathrm{prompt}} < 8.5\,\mathrm{MeV}$ and 6351.80 days for $E_{\mathrm{prompt}} \geq 8.5\,\mathrm{MeV}$. The two livetimes differ because the dead time introduced by the shower-likelihood cut described in Section~\ref{sec:background} depends on energy. After applying all conventional antineutrino selections, including the delayed-coincidence, fiducial-volume, and shower-likelihood selections, 28 candidate events remain. Applying the additional KamNet-based selection reduces the final sample to seven candidates. The consistency between the KamNet selection efficiency estimated from simulations and that inferred from the candidate sample is validated using toy MC studies; the observed number of accepted candidates is consistent with our expectation. Figure~\ref{fig:property} shows the distributions of these events before and after the final KamNet-based selection.
Figure~\ref{fig:property_vertex} shows the event positions for the final DSNB candidates.

The expected DSNB spectrum depends on both the cosmic core-collapse supernova rate and the neutrino emission from individual supernovae. Consequently, the signal shape entering the spectral fit is model dependent. To account for this dependence, we repeat the spectral analysis for several representative DSNB models spanning the historical development of DSNB calculations and a range of astrophysical assumptions. These include early phenomenological models, models based on observationally constrained star-formation histories, and more recent calculations incorporating black-hole-forming collapses, metallicity evolution, and variations of the stellar initial mass function.

Although the observed event count lies below the nominal background expectation, it remains statistically consistent with the background-only hypothesis once the sizable systematic uncertainties are taken into account. We perform a spectral analysis using the prompt-event energy and reconstructed radial distributions. In energy, the DSNB signal is expected to peak around 10\,MeV, whereas reactor antineutrinos and spallation products exhibit tails extending from the lower bound of the analysis window up to around 10\,MeV. The fast neutron is assumed to have a flat shape over the analysis energy region. In radius, fast neutrons are known to increase toward larger radii, while the other components are taken to be spatially uniform. The test statistic used in the spectral analysis is the same as in \cite{Abe_2022}, except that we employ a Poisson likelihood for the rate term.

\begin{figure}[htbp]
    \centering
    \includegraphics[width=0.8\linewidth]{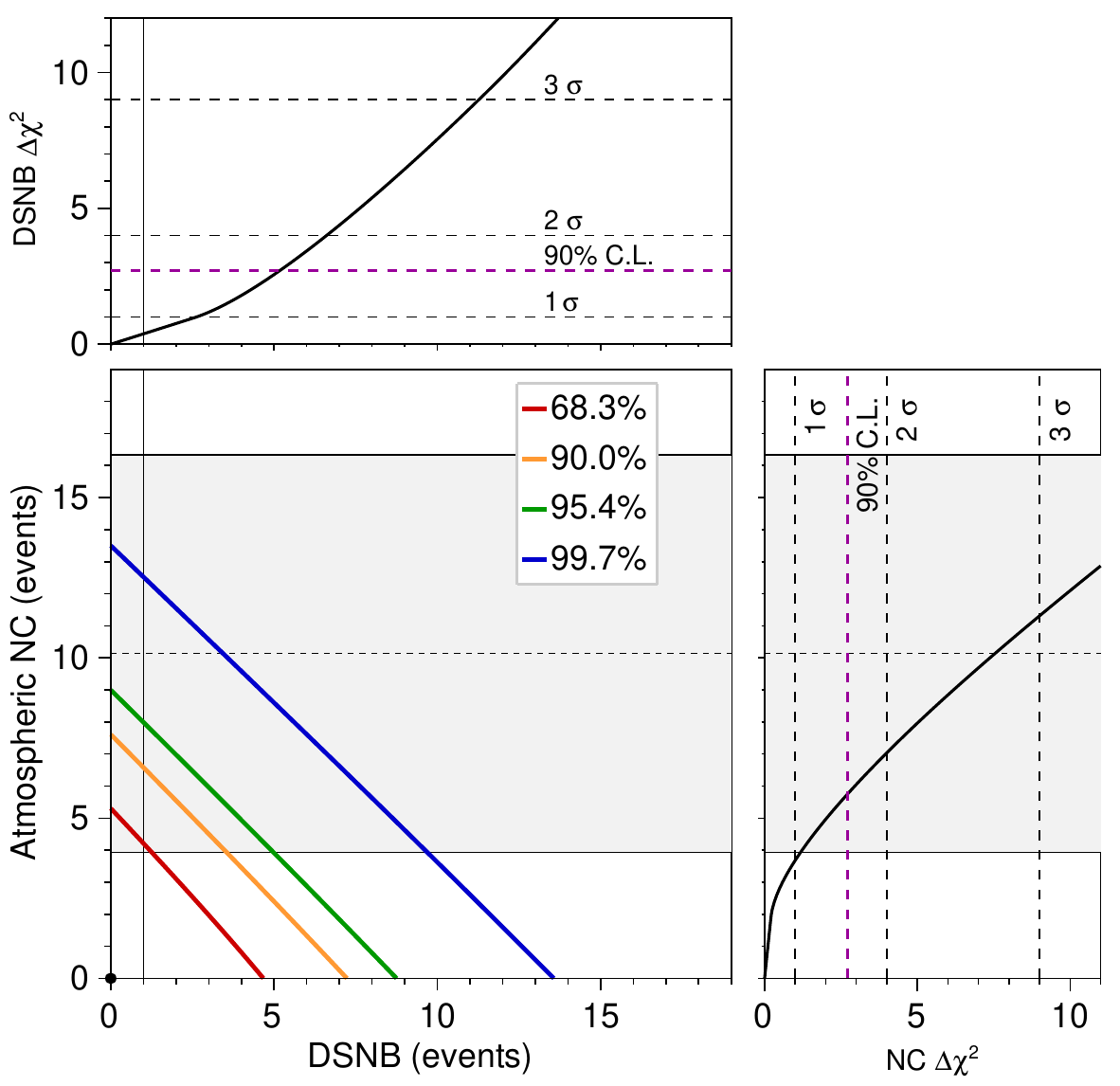}
    \caption{Two-dimensional scan of the number of DSNB and atmospheric-neutrino NC events. The DSNB model of \cite{PhysRevD.79.083013} is adopted. Color contours indicate the $1\sigma$ (red), 90\% (orange), $2\sigma$ (green), and $3\sigma$ (blue) allowed regions. The best-fit numbers of DSNB and NC events are both 0.0 (black circle). The horizontally hatched region denotes the expected number of NC events with its $1\sigma$ uncertainty, while the vertical line indicates the expected number of DSNB events. The top and right panels show the one-dimensional $\Delta \chi^2$ distributions for the DSNB and NC event components, respectively. The 90\% CL upper limit on the number of DSNB events is 5.2.}
    \label{fig:contour}
\end{figure}

\begin{figure}[htbp]
\centering
    \begin{minipage}{0.49\columnwidth}
        \centering
        \includegraphics[page=1, width=8.5cm]{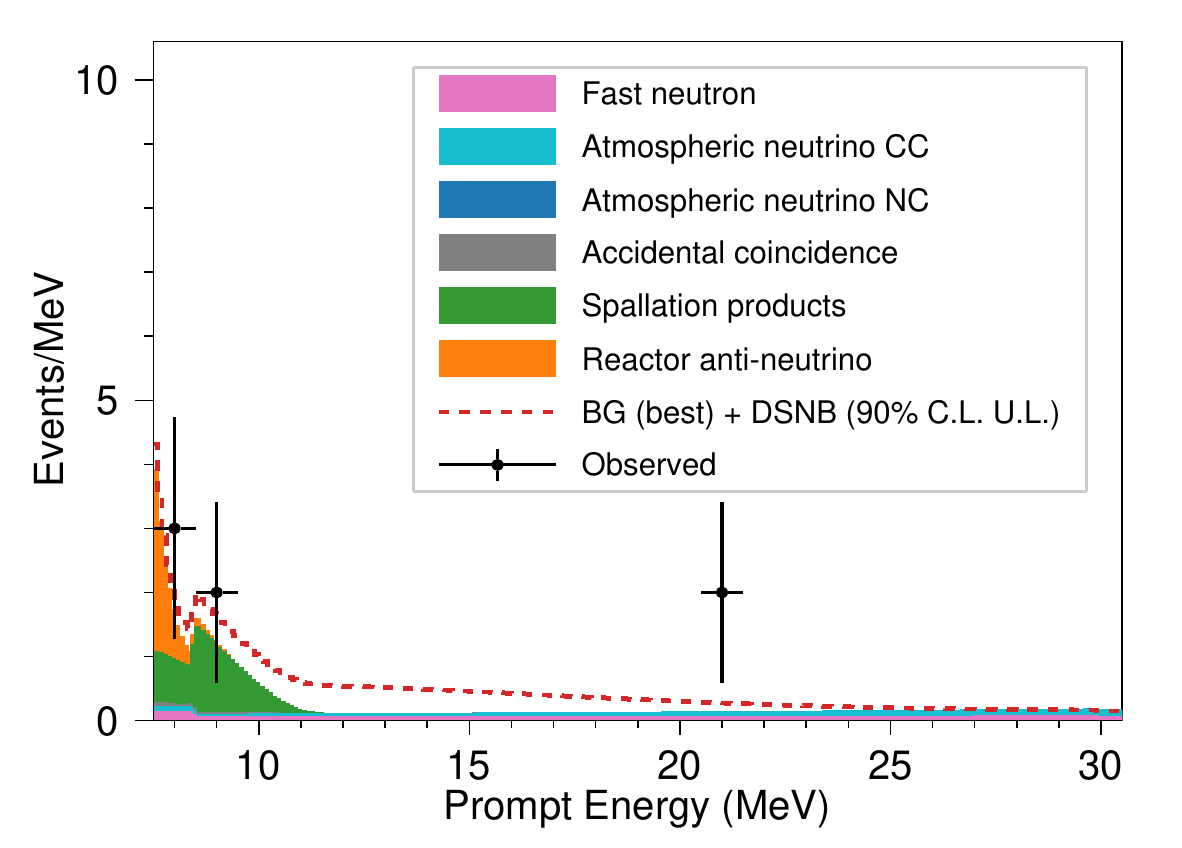}
    \end{minipage}
    \begin{minipage}{0.49\columnwidth}
        \centering
        \includegraphics[page=2, width=8.5cm]{Vertex_Horiuchi.pdf}
    \end{minipage}
    \caption{Prompt energy spectrum and radial distribution of the best-fit backgrounds and the DSNB signal at the 90\% C.L. upper limit. The DSNB model of \cite{PhysRevD.79.083013} is adopted. All histograms are stacked.}
    \label{fig:energy_vertex}
\end{figure}

In the fit, the reactor, spallation, and fast-neutron backgrounds are constrained by their estimated values and uncertainties, the atmospheric-neutrino CC component is left free to float, and the accidental background is fixed because of its negligible contribution. Because the DSNB and atmospheric-neutrino NC are correlated in the fit, we perform a two-dimensional scan over the number of events of these two components. Figure~\ref{fig:contour} shows the result of the two-dimensional scan for the DSNB signal (Horiuchi model) and the atmospheric-neutrino NC events. The expected and best-fit number of events for each component are summarized in Table~\ref{tab:exp_bestfit}. In this case, the best-fit number of DSNB events is zero, with a 90\% CL upper limit of 5.2 events. The best-fit number of atmospheric-neutrino NC events is also zero; however, the 1$\sigma$ allowed region overlaps with the uncertainty on the expected events, indicating no significant tension with the prediction. Figure~\ref{fig:energy_vertex} shows the prompt-energy and radial distributions for the best-fit background model and for the DSNB contribution set to its 90\% CL upper limit. The step at 8.5\,MeV in the energy distribution arises because the selection efficiencies of the KamNet-based cut and the shower-likelihood cut change across this energy boundary. Compared to the NC component, the atmospheric CC component prefers a higher-energy spectrum. In our dataset, two events are observed near 20\,MeV; this drives the inversion between the NC and CC contributions when comparing the expected events to the best-fit events.

If the KamNet selection efficiency varies with energy within 8.5--30\,MeV according to its systematic uncertainty, the resulting distortion can modify the inferred spectral shapes and thus affect the fit. To evaluate this impact, we construct multiple distortion patterns for the signal and background spectra consistent with the systematic uncertainty and repeat the spectral analysis for combinations of these patterns. From the distribution of the resulting upper limits, we obtain the standard deviation and conservatively shift the nominal upper limit by this amount to derive the final upper limit on the number of DSNB signals. The 90\% CL upper limit on the DSNB flux, $F_{90}$, is obtained from the corresponding upper limit on the number of DSNB events, $N_{90}$, using
\begin{equation}
    F_{90} = N_{90} \times \frac{F_{\mathrm{exp}}}{N_{\mathrm{exp}}},
\end{equation}
where $F_{\mathrm{exp}}$ and $N_{\mathrm{exp}}$ are the predicted DSNB flux and the corresponding expected number of events for a given model, respectively. Table~\ref{tab:DSNBflux} summarizes the predicted fluxes for several DSNB models and the corresponding upper limits obtained in this work. The selected models span the historical development of DSNB predictions over the past three decades. \citet{TOTANI1995367} presented an early calculation exploring the dependence of the SRN spectrum on cosmological parameters, while \citet{PhysRevD.62.043001} derived an astrophysically constrained upper bound based on the observed cosmic metal-enrichment history. \citet{PhysRevD.79.083013} subsequently established a widely used framework based on an observationally constrained cosmic star-formation history. \citet{Nakazato_2013,Nakazato_2015} incorporated metallicity-dependent black-hole formation, while \citet{Ashida_2023} explored a galactic chemical-evolution model with a galaxy-dependent initial mass function. These models predict integrated DSNB fluxes above 8.3~MeV ranging from approximately 1.4 to 25.0~$\mathrm{cm^{-2}\,s^{-1}}$, as summarized in Table~\ref{tab:DSNBflux}. These predictions do not encompass all sources of astrophysical uncertainty. In particular, recent work by \citet{Anandagoda_2023} quantified the impact of different cosmic star-formation-rate-density models on the predicted DSNB flux, demonstrating that uncertainties in the cosmic star-formation history remain an important source of uncertainty in the predicted DSNB flux. 

Although no significant signal excess is observed for any of the DSNB models, we obtained 90\% CL upper limits on the DSNB flux of 38--43\,$\mathrm{cm^{-2}\,s^{-1}}$. This result represents a substantial improvement over the previous KamLAND result (60--110\,$\mathrm{cm^{-2}\,s^{-1}}$) and demonstrates that the implementation of KamNet-based event selection provides an improvement beyond that expected from the increased exposure alone.

\begin{table}[hbtp]
    \caption{Summary of estimated background and best-fit parameters}
    \label{tab:exp_bestfit}
    \centering
    \begin{tabular}{lcc}
        \toprule
        Source & Expected & Best fit \\
        \midrule
        \midrule
        Reactor $\bar{\nu}_e$ & $1.0 \pm 0.6$ & 1.0 \\
        Spallation $^{9}\mathrm{Li}$ & $1.9 \pm 6.6$ & 2.3 \\
        Atmospheric neutrino CC & $0.5 \pm 0.3$ & 1.6 \\
        Atmospheric neutrino NC & $10.1 \pm 6.2$ & 0.0 \\
        Fast neutron & $2.5 \pm 2.5$ & 1.2 \\
        Accidental background & $0.11 \pm 0.02$ & 0.11 \\
        \midrule
        Total (background only) & $16.2 \pm 9.4$ & 6.2 \\
        \midrule
        DSNB (Horiuchi+09) & 1.0 & 0.0 \\
        \midrule
        Observed & 7 & \\
        \bottomrule
    \end{tabular}
\end{table}

\begin{table}[hbtp]
    \caption{Summary of fitting results and predicted values for the DSNB fluxes (in $\mathrm{cm^{-2} \, s^{-1}}$)}
    \label{tab:DSNBflux}
    \centering
    \begin{tabular}{lccccc}
        \toprule
        DSNB model & Reference & Best fit & 90\% C.L. U.L. & Predicted & Role \\
        \midrule
        \midrule
        Totani+95 (constant) & \cite{TOTANI1995367} & $0.00^{+24.9}_{-0.0}$ & 38.6 & 25.0 & early cosmological model \\
        Kaplinghat+00 & \cite{PhysRevD.62.043001} & $0.00^{+19.8}_{-0.0}$ & 37.9 & 17.9 & metal-enrichment bound \\
        Horiuchi+09 (6\,MeV) & \cite{PhysRevD.79.083013} & $0.00^{+23.7}_{-0.0}$ & 39.1 & 6.7 & observational SFR \\
        Nakazato+15 (max, IH) & \cite{Nakazato_2013, Nakazato_2015} & $0.00^{+20.7}_{-0.0}$ & 40.3 & 3.3 & BH formation + metallicity \\
        Nakazato+15 (min, NH) & \cite{Nakazato_2013, Nakazato_2015} & $0.00^{+19.5}_{-0.0}$ & 38.6 & 1.4 & BH formation + metallicity  \\
        Ashida+23 (HB06, NH) & \cite{Ashida_2023} & $0.00^{+24.3}_{-0.0}$ & 43.0 & 12.2  & 
        galaxy-dependent IMF \\
        \bottomrule
    \end{tabular}
\end{table}

\begin{figure}[htbp]
    \centering
    \includegraphics[width=0.9\linewidth]{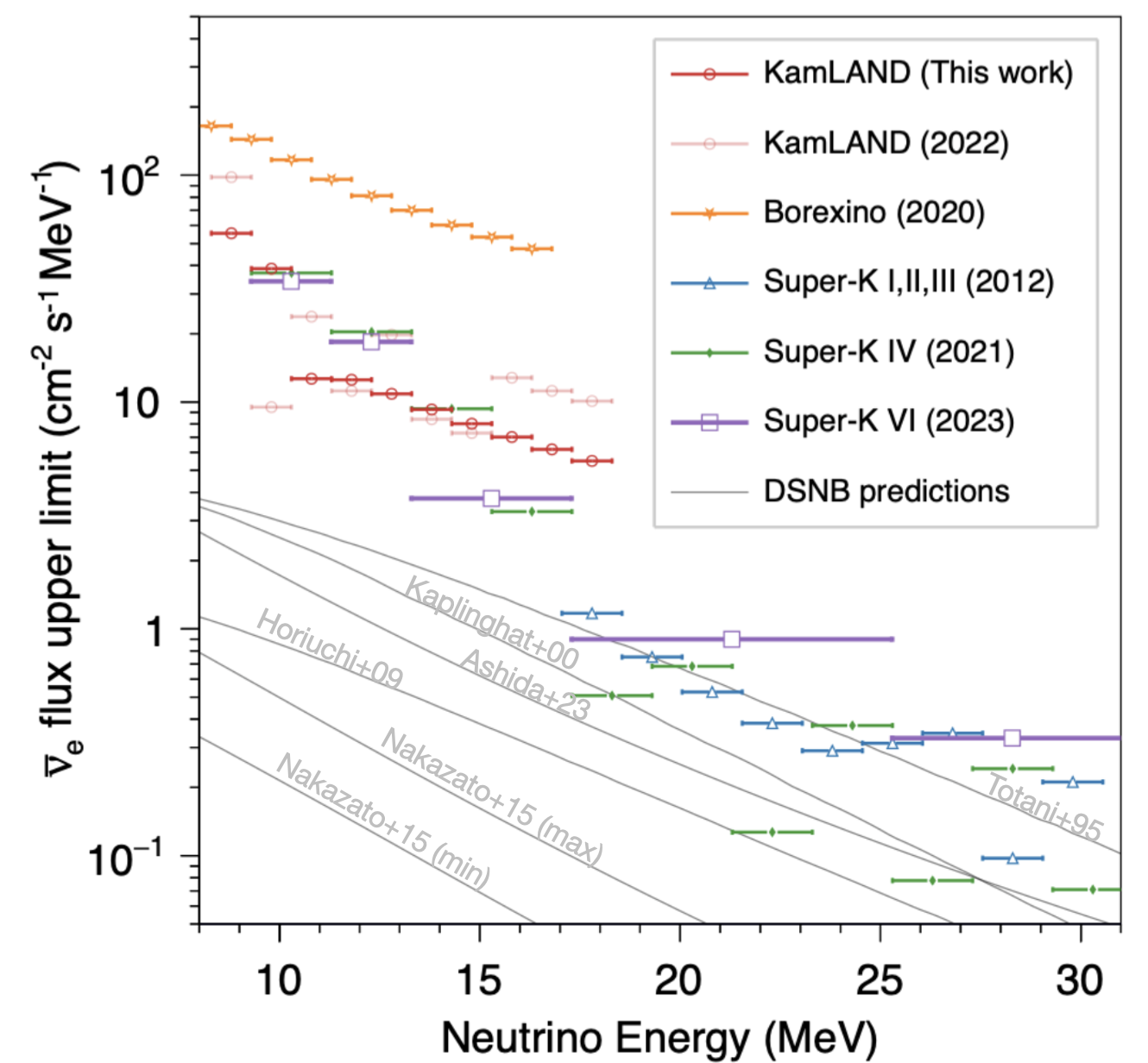}
    \caption{The 90\% C.L. upper limits on the model-independent $\bar{\nu}_e$ flux. The red points represent the limits obtained in this work, while the lighter red points indicate the limits from the previous KamLAND result~\citep{Abe_2022}. Other points show the limits reported by Borexino~\citep{AGOSTINI2021102509} and Super-Kamiokande during the SK-I, II, III~\citep{PhysRevD.85.052007}, SK-IV~\citep{PhysRevD.104.122002}, and SK-VI~\citep{Harada_2023} periods. The lighter black lines correspond to the DSNB prediction models.}
    \label{fig:model_independent}
\end{figure}

We also present model-independent upper limits on the $\bar{\nu}_e$ flux. Assuming a monochromatic $\bar{\nu}_e$ source, the 90\% CL upper limit on the flux in each energy bin can be written as
\begin{equation}
    \phi_{90} = \frac{N_{90}}{N_p \cdot \sigma \cdot \epsilon_{\mathrm{IBD}} \cdot T},
\end{equation}
where $N_{90}$ is the 90\% CL upper limit on the number of $\bar{\nu}_e$ events in the bin, determined from the observed and best-fit background counts using the Feldman-Cousins method~\citep{PhysRevD.57.3873}; $N_p$ is the number of target protons; $\sigma$ is the IBD cross section; $\epsilon_{\mathrm{IBD}}$ is the detector efficiency; and $T$ is the livetime. Figure~\ref{fig:model_independent} shows the resulting model-independent 90\% CL upper limits on the $\bar{\nu}_e$ flux, together with the previous KamLAND result~\citep{Abe_2022}, results from other experiments (Borexino~\citep{AGOSTINI2021102509} and Super-Kamiokande~\citep{PhysRevD.85.052007, PhysRevD.104.122002, Harada_2023}), and representative DSNB model predictions. Table~\ref{tab:model_independent} lists the model-independent 90\% CL upper limits derived in this work. Because the KamLAND sample is statistics-limited, the appearance of a single event can weaken the limit in a given energy bin compared to the 2022 result ($E_\nu = 9.3\text{--}10.3$\,MeV). In addition, because zero events are observed in both the 2022 KamLAND study and this work, the Feldman-Cousins upper limit is affected by the expected background level. Due to the background reduction by the KamNet selection, a smaller expected background can lead to a less stringent upper limit ($E_\nu = 11.3\text{--}12.3$, 13.3--14.3, 14.3--15.3\,MeV). While Super-Kamiokande provides stronger limits at higher energies because of its much larger exposure, KamLAND remains complementary in the lower-energy region and provides particularly strong constraints below 13.3\,MeV.

\begin{table}[hbtp]
    \caption{90\% C.L. upper limits on the model-independent $\bar{\nu}_e$ flux}
    \label{tab:model_independent}
    \centering
    \begin{tabular}{cc}
        \toprule
        Neutrino energy & Flux 90\% C.L. upper limit \\
        $[\mathrm{MeV}]$ & [$\mathrm{cm^{-2}\,s^{-1}\,MeV^{-1}}$] \\
        \midrule
        8.3--9.3 & 55.5 \\
        9.3--10.3 & 38.7 \\
        10.3--11.3 & 12.7 \\
        11.3--12.3 & 12.5 \\
        12.3--13.3 & 10.9 \\
        13.3--14.3 & 9.3 \\
        14.3--15.3 & 8.0 \\
        15.3--16.3 & 7.0 \\
        16.3--17.3 & 6.2 \\
        17.3--18.3 & 5.5 \\
        \bottomrule
    \end{tabular}
\end{table}

\section{Discussion} \label{sec:discussion}

\begin{figure}[htbp]
    \centering
    \includegraphics[width=0.8\linewidth]{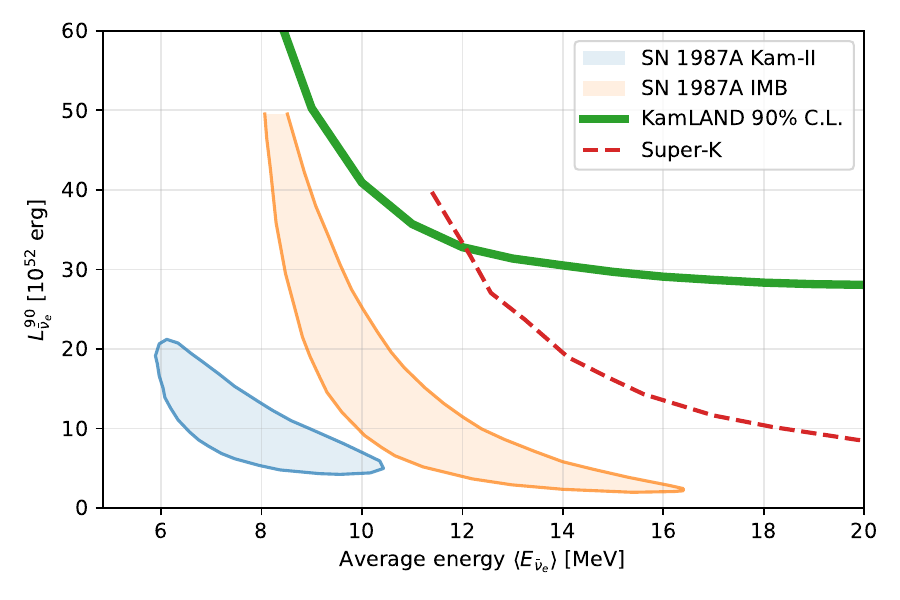}
    \caption{Upper limits (90\% C.L.) on the total emitted $\bar{\nu}_e$ energy, $L_{\bar{\nu}_e}$, as a function of the average neutrino energy, $\langle E_{\bar{\nu}_e}\rangle$.  The Super-Kamiokande constraint~\citep{PhysRevD.104.122002} and the 68.3\% C.L. regions inferred from the SN 1987A neutrino observations~\citep{Mirizzi2005}, assuming $\alpha=2$, are overlaid for comparison. }
    \label{fig:lumi_upper}
\end{figure}

To investigate the astrophysical implications of the present KamLAND results, we repeat the full analysis for a series of pinched thermal neutrino spectra with different average electron-antineutrino energies, $\langle E_{\bar{\nu}_e}\rangle$. The neutrino spectrum is parameterized following \cite{Keil2003} with a pinching parameter of $\alpha=2$. This choice also facilitates a direct comparison with the SN~1987A constraints of Mirizzi and Raffelt~\citep{Mirizzi2005}, whose Figure 2 was obtained assuming the same value $\alpha=2$. We fix the normalization and redshift evolution of the cosmic core-collapse supernova rate using the observationally constrained cosmic SFR adopted by Horiuchi, Beacom, and Dwek~\citep{PhysRevD.79.083013}. This SFR model was cross-checked against measurements of the supernova rate, extragalactic background light, and stellar mass density. The resulting 90\% C.L. upper limits on the total emitted electron-antineutrino energy, $L_{\bar{\nu}e}$, are shown in Figure~\ref{fig:lumi_upper} as a function of $\langle E{\bar{\nu}_e}\rangle$.

For comparison, we also show the corresponding constraint derived from the combined Super-Kamiokande Phase I--IV analysis~\citep{PhysRevD.104.122002} and the 68.3\% C.L. allowed regions inferred from the SN~1987A neutrino observations reported by Mirizzi and Raffelt~\citep{Mirizzi2005}. The latter were obtained assuming $\alpha=2$, consistent with the spectral-shape parameter adopted in the present analysis. Presenting these results in the same $L_{\bar{\nu}_e}$--$\langle E_{\bar{\nu}_e}\rangle$ parameter space enables a direct comparison between the present DSNB constraint and the neutrino emission inferred from SN~1987A.

Figure~\ref{fig:lumi_upper} demonstrates the complementary sensitivities of KamLAND and Super-Kamiokande. Owing to its low analysis threshold and effective background rejection, KamLAND provides the most stringent constraints for relatively soft neutrino spectra with $\langle E_{\bar{\nu}_e}\rangle \lesssim 12$~MeV, whereas the substantially larger fiducial volume of Super-Kamiokande leads to stronger constraints at higher average neutrino energies. Although the current KamLAND sensitivity does not yet reach the parameter region favored by the SN~1987A observations, the figure provides an intuitive astrophysical interpretation of the present DSNB limits and illustrates how liquid-scintillator and water-Cherenkov detectors probe complementary regions of the core-collapse supernova neutrino-emission parameter space. Future improvements in detector exposure and background suppression, together with measurements by multiple detector technologies, will further strengthen constraints on the average neutrino emission from core-collapse supernovae.

\section{Summary} \label{sec:summary}
We performed a search for the DSNB using the full KamLAND dataset, corresponding to more than 6000 days of livetime collected from March 2002 to August 2024. To classify backgrounds accompanied by proton recoils from neutrons, we adopted a deep-neural-network-based model, KamNet, and achieved a background rejection efficiency of up to 91.3\%. The search covered the neutrino energy range of 8.3--30.8\,MeV, and we observed seven candidate events. This was consistent with the background-only expectation of $16.2 \pm 9.4$ events, which included the effects of the KamNet selection efficiency and its associated systematic uncertainty. Using a spectral analysis based on the prompt-energy and radial distributions, we found that the best-fit number of DSNB events was zero for all DSNB models considered. We therefore set 90\% CL upper limits on the DSNB flux, obtaining values of 38--43\,$\mathrm{cm^{-2}\,s^{-1}}$.  We also present model-independent upper limits on the $\bar{\nu}_e$ flux, which are among the most stringent constraints below 13.3\,MeV. Future KamLAND DSNB searches could further improve their sensitivity by combining the KamNet-based selection with recently developed spallation-rejection techniques targeting cosmogenic isotope backgrounds~\citep{2510.12873}. 

More broadly, this work demonstrates that spatiotemporal PMT-hit information can be used to suppress proton-recoil-associated backgrounds in liquid-scintillator detectors, an approach that is relevant for future searches in KamLAND2, JUNO~\citep{2022103927}, and THEIA~\citep{2020EPJC...80..416A}. Together with complementary measurements by water Cherenkov detectors, these experiments will provide increasingly powerful probes of the average neutrino emission from core-collapse supernovae.


\begin{acknowledgments}
The KamLAND experiment is supported by 
JSPS KAKENHI grants 
JP24H02237; 
the World Premier International Research Center Initiative (WPI Initiative), MEXT, Japan; 
Netherlands Organization for Scientific Research (NWO); 
the Heising-Simons Foundation;
and under the U.S. Department of Energy (DOE) Contract 
No.~DE-AC02-05CH11231,
the National Science Foundation (NSF) 
NSF-2012964, 
NSF-2110720, 
as well as other DOE and NSF grants to individual institutions.  
The Kamioka Mining and Smelting Company has provided services for activities in the mine.  
We acknowledge the support of NII for SINET. 
This work is partly supported by 
the Graduate Program on Physics for the Universe (GP-PU).
\end{acknowledgments}


\bibliography{main.bib}{}
\bibliographystyle{aasjournal}

\end{document}